\documentclass[12pt,a4paper]{article}
\usepackage{epsfig}
\newcommand{\gs}   {$\gamma_{\rm s}\ $}
\newcommand{\ppb}  {$\rm{p\bar{p}}\;$}
\newcommand{\ssb}  {$\rm{s\bar{s}}\;$}
\newcommand{\pp}   {$\rm{pp}$ }
\newcommand{\ls}   {$\lambda_{\rm s}\ $}
\newcommand{\qi}   {{\mathbf{q}}_i}
\newcommand{\vexp} {$VT^3\exp{(-0.7 {\rm GeV}/T)}$}
\newcommand{\muv}  {{\mbox{\boldmath ${\mu}\!$ \unboldmath} }}
\begin{document}
\begin{center}
\Large{\bf Common Features of Particle 
Multiplicities 
in  Heavy Ion Collisions.}
\end{center}
\begin{center}
\author{F. Becattini}
{\it Universit\`a di Firenze and INFN Sezione di Firenze,\\ 
Largo E. Fermi 2, I-50125, Florence, Italy}\\
\author{J. Cleymans}
{\it Department of Physics, University of Cape Town,\\
Rondebosch 7701, Cape Town, South Africa}\\
\author{A. Ker\"anen, E. Suhonen}
{\it Department of Physical Sciences, University of Oulu,\\
FIN-90571 Oulu, Finland}\\
\author{K. Redlich}
{\it Institute of Theoretical Physics, University of Wroclaw,\\
PL-50204 Wroclaw, Poland.}\\
\end{center}
\begin{abstract}
Results of a systematic study 
of fully integrated particle multiplicities
in central Au--Au and Pb--Pb collisions at beam momenta of 1.7$A$ GeV, 11.6$A$ GeV 
(Au--Au) and 158$A$ GeV (Pb--Pb)  using a statistical-thermal model are presented.
The close similarity of the colliding systems makes it possible to study
heavy ion collisions under definite initial conditions over a range of centre-of-mass 
energies covering more than one order of magnitude. 
We conclude that a thermal model description of particle 
multiplicities, with additional strangeness suppression, is possible for each energy.
The degree of chemical equilibrium of strange particles and the relative production 
of strange quarks with respect to u and d quarks are higher than in
 {${\rm e}^+{\rm e}^-$}, \pp and 
\ppb collisions at comparable and even at lower energies. 
The average energy per hadron in the 
comoving frame is always close to 1 GeV per hadron
 despite the fact that the energy varies
more than 10-fold.
\end{abstract}
\section{Introduction}
It is becoming more and more clear that 
results from relativistic heavy ion collisions at many different
energies\cite{heinz} show striking  common traits.
Statistical-thermal models are able to 
reproduce particle multiplicities in a satisfactory manner by using 
a very small number of 
parameters: temperature, volume, baryon chemical potential and a possible 
strange-quark suppression parameter, \gs \cite{gs}. 
We report here the results\cite{bcksr} of
an analysis
of data from  
collisions
at several different energies, with
 emphasis on the similarity of the colliding system.
We have focussed our attention on central Au--Au collisions at 
beam momenta of 1.7$A$ GeV (SIS) \cite{oeschler}, 11.6$A$ GeV (AGS) \cite{ogilvie}
and on central Pb--Pb collisions at 158$A$ GeV (SPS) beam momentum \cite{stock}.
As far as the choice of data (and, consequently, colliding system) is concerned, 
our leading rule is the availability of full phase space integrated multiplicity 
measurements because a pure statistical-thermal model analysis of particle yields, 
without any consideration of dynamical effects, {\em may} apply only in this case 
\cite{prc}. Such data, however, exist only in a few cases and whenever legitimate 
we have extrapolated spectra measured in a limited rapidity window to full phase
space. The use of extrapolations is more correct than using data over limited 
intervals of rapidity, especially in the framework of a purely statistical-thermal 
analysis without a dynamical model. Moreover, the usually employed requirement 
of zero strangeness ($S=0$) demands fully integrated multiplicities because 
strangeness does not need to vanish in a limited region of phase space.\\

In order to assess the consistency of the results obtained, we have performed the
statistical-thermal model analysis by using two completely independent numerical
algorithms whose outcomes turned out to be in close agreement throughout.  
Similar analyses have been recently made by other authors (see e.g. \cite{heppe,letessier});
however, both the model and the used data set differ in several important details, 
such as the assumption of full or partial equilibrium for some quark flavours, 
the number of included resonances, the treatment of resonance widths, inclusion 
or not of excluded volume corrections, treatment of flow, corrections due to limited 
rapidity windows etc. Because of these differences it is  difficult 
to trace the origin of discrepancies between different results. We hope that the 
present analysis, covering a wide range of beam energies using a consistent 
treatment, will make it easier to appreciate the energy dependence of the various 
parameters such as temperature and chemical potential.

\section{Data set and model description}

As emphasized in the introduction, in the present analysis we use the most recent 
available data, concentrating on fully integrated particle yields and discarding 
data that have been obtained in limited kinematic windows.
We have derived integrated multiplicities of $\mathrm{\pi}^+$, $\Lambda$ and proton 
in Au--Au collisions at AGS by extrapolating published rapidity distributions 
\cite{piAu,LAu,pAu} with constrained mid-rapidity value ($y_{\rm NN}$=1.6). For
proton and $\Lambda$ we have fitted the data to Gaussian distributions, whilst for
$\pi^+$ we have used a symmetric flat distribution at midrapidity with Gaussian-shaped 
wings on either side; the point at which the Gaussian wing and the plateau connect
is a free parameter of the fit. The fits yielded very good $\chi^2$'s/dof: 0.27, 1.24 
and 1.00 for $\pi^+$, proton and $\Lambda$ respectively. The integrated multiplicities 
have been taken as the area under the fitted distribution between the 
minimal $y_{\rm min}$ and maximal $y_{\rm max}$ values of rapidities for the reactions 
${\rm N N} \rightarrow \pi {\rm N N}$, ${\rm N N} \rightarrow \Lambda {\rm K}$ for 
pions and $\Lambda$'s respectively; the difference between these areas and the total 
area has been taken as an additional systematic error. The area between $y_{\rm min}$ 
and $y_{\rm max}$ amounts to practically 100\% of the total area for pions and about 
95\% for $\Lambda$'s.

We have not included data on deuteron production because of the possible inclusion 
of fragments in the measured yields. This is particularly dangerous at low (SIS) 
energies where inclusion or not of deuterons modifies thermodynamic quantities like 
$\epsilon /n$ \cite{prl}.\\  
The data analysis has been performed within an ideal hadron gas grand-canonical 
framework supplemented with strange quark fugacity \gs. In this approach, 
the overall average multiplicities of hadrons and hadronic resonances are determined 
by an integral over a statistical distribution:       

\begin{equation}
\langle n_i \rangle = (2 J_i + 1)\frac{V}{(2\pi)^3} \int {\rm d}^3p \,\,
\frac{1}{\gamma_{\rm s}^{-s_i} \exp{[(E_i-\muv \cdot \qi)/T]} \pm 1}
\label{eqn:thermaldist}
\end{equation}
where $\qi$ is a three-dimensional vector with electric charge, baryon number and 
strangeness of hadron $i$ as components; \muv the vector of relevant chemical 
potentials; $J_i$ the spin of hadron $i$ and $s_i$ the number of valence strange 
quarks in it; the $+$ sign in the denominator is relevant for fermions, the $-$ 
for bosons. This formula holds in case of many different statistical-thermal 
systems (i.e. clusters or fireballs) having common temperature and \gs but  
different arbitrary momenta, 
provided that the probability of realizing a given distribution of quantum numbers 
among them follows a statistical rule \cite{bgs,beca}. In this case $V$ must be 
understood as the sum of all cluster volumes measured in their own rest frame. 
Furthermore, since both volume and participant nucleons may fluctuate on an event 
by event basis, $V$ and \muv (and maybe $T$) in Eq.~(\ref{eqn:thermaldist}) should be 
considered as average quantities \cite{bgs}.\\ 
The overall abundance of a hadron of type $i$ to be compared with experimental 
data is determined by the sum of Eq.~(\ref{eqn:thermaldist}) and the contribution 
from decays of heavier hadrons and resonances:

\begin{equation}
  n_i = n_i^{\rm primary} + \sum_j {\rm Br}(j\rightarrow i) n_j
\label{eqn:sum}
\end{equation}
where the branching ratios Br$(j\rightarrow i)$ have been taken from the 1998
issue of the Particle Data Table \cite{PDG}.\\ 
It must be stressed that the unstable hadrons contributing to the sum in 
Eq.~(\ref{eqn:sum}) may differ according to the particular experimental definition. 
This is a major point in the analysis procedure because quoted experimental 
multiplicities may or may not include contributions from weak decays of 
hyperons and K$^0_S$. We have included all weak decay products in our computed 
multiplicities except in Pb--Pb collisions on the basis of relevant statements
in ref.~\cite{sikler} and about antiproton production in refs.~\cite{Si3,antipAu}. 
It must be noted that switching this assumption in Au--Au at SIS and AGS does not 
affect significantly the resulting fit parameters.\\
The overall multiplicities of hadrons depend on several unknown parameters
(see Eq.~(\ref{eqn:thermaldist})) which are determined by a fit to the data. 
The free parameters in the fit are $T$, $V$, \gs and $\mu_B$ (the baryon 
chemical potential) whereas $\mu_S$ and $\mu_Q$, i.e. the strangeness and electric 
chemical potentials, are determined by using the constraint of overall 
vanishing strangeness and forcing the ratio between net electric charge and
net baryon number $Q/B$ to be equal to the ratio between participant protons
and nucleons. The latter is assumed to be $Z/A$ of the colliding nucleus in 
Au--Au and Pb--Pb.\\
For SIS Au-Au data we have required the exact conservation of 
strangeness instead of using a strangeness chemical potential. This gives rise
to slightly more complex calculations which are necessary owing
to the very small strange particle production (Au--Au).
The difference between these strangeness-canonical 
and pure grand-canonical calculations of multiplicities of K and 
$\Lambda$ for the final set of thermal parameters (see Table 1) turns out to 
be as large as a factor 
15 in Au--Au at 1.7$A$ GeV.\\
Owing to few available data points in SIS Au--Au collisions, we have not fitted 
the volume $V$ nor the \gs  therein. The volume has been assumed to be 
$4\pi r^3/3$ where $r = 7$ fm (approximately the radius of a Au nucleus) while 
\gs has been set to 1, the expected value for a completely equilibrated hadron 
gas. Since we have performed 
a strangeness-canonical calculation here, the yield 
ratios involving strange particle are not independent of the chosen volume value
as in the grand-canonical framework. Thus, in this particular case, $V$ is 
meant to be the volume within which strangeness is conserved (i.e. vanishing) 
and not the global volume defining overall particle multiplicities as in 
Eq.~(\ref{eqn:thermaldist}). Also, in order to test the dependence of this 
assumption on our results, we have repeated the fit by varying $V$ by a factor 2 
and 0.5 in turn.\\     
A major problem in Eq.~(\ref{eqn:sum}) is where to stop the summation over 
hadronic states. Indeed, as mass increases, our knowledge of the hadronic 
spectrum becomes less accurate; starting from $\approx 1.7$ GeV many states 
are possibly missing, masses and widths are not well determined and so are the 
branching ratios. For this reason, it is unavoidable that a cut-off on hadronic 
states be introduced in Eq.~(\ref{eqn:sum}). If the calculations are sensitive 
to the value of this cut-off, then the reliability of results is questionable. 
We have performed all our calculations with two cut-offs, one at around 1.8 GeV 
(in the analysis algorithm A) and the other one at 2.4 GeV (in the analysis 
algorithm B). The contribution of missing heavy resonances is expected to be 
very important for temperatures $\geq 200$ MeV making thermal models inherently 
unreliable above this temperature.
\begin{small}
\begin{table}[htb]
\begin{center}
\caption{Summary of fit results. Free fit parameters are quoted along with 
resulting minimum $\chi^2$'s and \ls parameters.}
\vspace{0.5cm}
\begin{tabular}{|c|c|} 
\hline
                        &  Average \\
\hline
\multicolumn{2}{|c|}{Au--Au 1.7$A$ GeV} \\
\hline                
$T$ (MeV)        &  49.6$\pm$2.5      \\
$\mu_B$ (MeV)    &  813$\pm$23        \\
\gs              &  1 (fixed)         \\
$V$(fm$^3$)      &  1437 (fixed)      \\
\ls              &  0.0054$\pm$0.0035 \\
\hline
\multicolumn{2}{|c|}{Au--Au 11.6$A$ GeV} \\
\hline                 
$T$ (MeV)          & 119.8$\pm$8.3     \\
$\mu_B$ (MeV)      & 553.5$\pm$16      \\
\gs                & 0.720$\pm$0.097   \\ 
\vexp              & 2.03$\pm$0.34     \\
\ls      &  0.43$\pm$0.10   \\                 
\hline
\multicolumn{2}{|c|}{Pb--Pb 158$A$ GeV} \\
\hline  
$T$ (MeV)     & 158.1$\pm$3.2    \\
$\mu_B$ (MeV) & 238$\pm$13        \\
\gs          & 0.789$\pm$0.052   \\
\vexp         & 21.7$\pm$2.6      \\
\ls          & 0.447$\pm$0.025   \\
\hline
\end{tabular}
\end{center}
\end{table}
\end{small}

\section{Results}

\begin{figure}[t]
\centerline{
\epsfxsize=25pc
\epsfbox{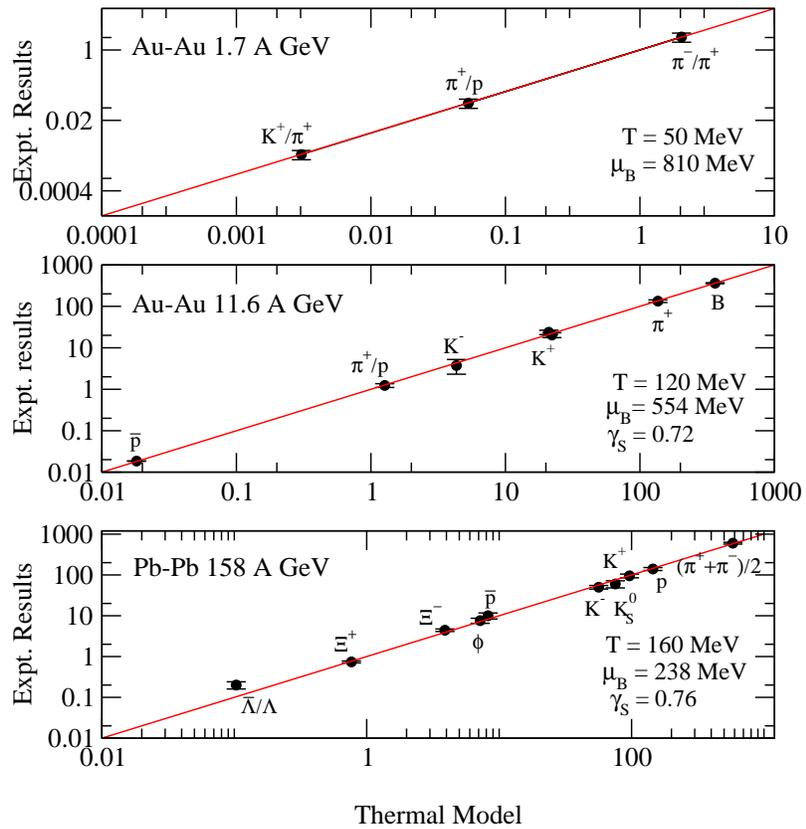}}
\caption{Comparison between  particle multiplicities fitted using
the thermal model and experimental results.}
\label{fig:big}
\end{figure}

As mentioned in the introduction, we have performed two analyses (A and B) 
by using completely independent algorithms.\\  
In the analysis A all light-flavoured resonances up to 1.8 GeV have been included. 
The production of neutral hadrons with a fraction $f$ of \ssb content has been 
suppressed by a factor $(1-f)+f\gamma_s^2$. In the analysis B the mass cut-off has 
been pushed to 2.4 GeV and neutral hadrons with a fraction $f$ of \ssb content 
have been suppressed by a factor $\gamma_s^{2f}$. Both algorithms use masses, widths 
and branching ratios of hadrons taken from the 1998 issue of Particle Data 
Table \cite{PDG}. However, it must be noted that differences between the two analyses 
exist in dealing with poorly known heavy resonance parameters, such as assumed 
central values of mass and width, where the Particle Data Table itself gives only 
a rough estimate. Moreover, the two analyses differ by the treatment of mass windows
within which the relativistic Breit-Wigner distribution is integrated.\\
A summary of the final results 
is shown in Fig.~\ref{fig:big}.
For each analysis an estimate of systematic errors on fit parameters have been
obtained by repeating the fit 
\begin{itemize}
\item assuming vanishing widths for all resonances 
\item varying the mass cut-off to 1.7 in analysis A and to 1.8 in 
analysis B
\item for Au--Au at 1.7$A$ GeV, the volume $V$ has been varied to $V/2$ 
and to $2V$ (see discussion in Sect.~2)   
\end{itemize}
The differences between new fitted parameters and main parameters have been 
conservatively taken as uncorrelated systematic errors to be added in quadrature 
for each variation (see Table 1).
 The effect of errors on masses, widths and branching 
ratios of inserted hadrons has been studied in analysis A according
 to the procedure 
described in ref.~\cite{bgs} and found to be negligible.\\
Finally, the results of the two analyses have been averaged according to a method 
suggested in ref.~\cite{schm}, well suited for strongly correlated measurements. 

\section{Discussion and conclusions}

From the results obtained, an indication emerges that a statistical-thermal 
description of multiplicities in a wide range of heavy ion collisions is indeed 
possible to a satisfactory degree of accuracy, for beam momenta ranging from 
1.7$A$ GeV to 158$A$ GeV per nucleon. Furthermore, the fitted parameters show a 
remarkably smooth and consistent dependence as a function of centre-of-mass
energy.

\begin{figure}[tbh]
\centerline{
\epsfxsize=25pc
\epsfbox{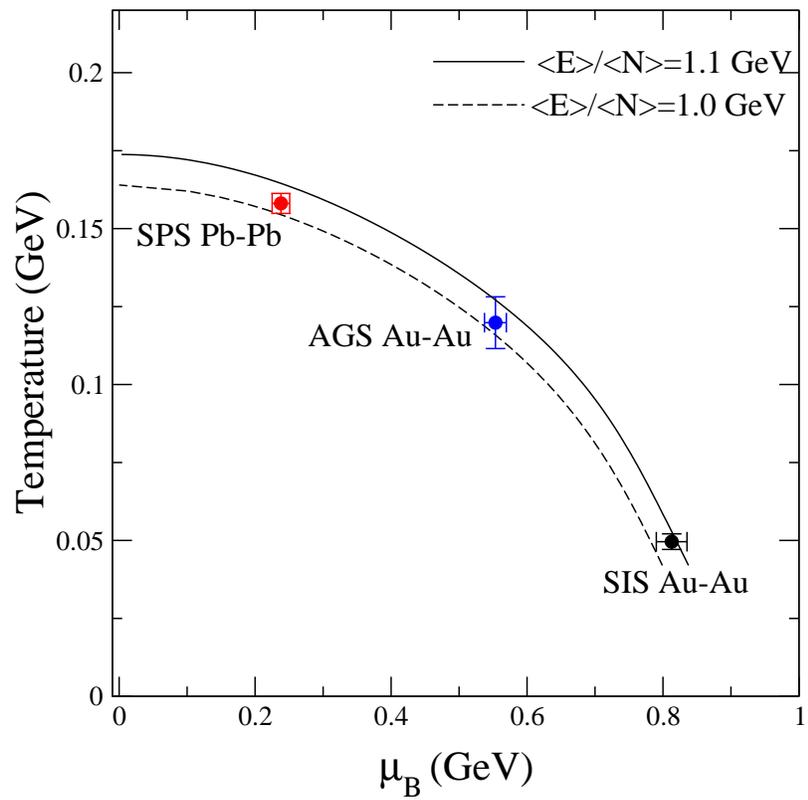}}
\caption{Fitted temperatures and baryon-chemical potentials plotted 
along with curves of constant energy per hadron.}
\label{fig:eovern}
\end{figure}

The temperature varies considerably between the lowest and the highest beam energy, 
namely, between 50 MeV at SIS and 160 MeV at SPS. Similarly, the baryon chemical 
potential changes appreciably, decreasing from about 820 MeV at SIS to about 240 
MeV at SPS. However, since the changes in temperature and chemical potential are
opposite, the resulting energy per particle shows little variation and remains 
practically constant at about 1 GeV per particle; this is shown in 
Fig.~\ref{fig:eovern}.\\
The supplementary \gs factor, measuring the deviation from a completely 
equilibrated hadron gas, is around 0.7 -- 0.8 at all energies where it has
been considered a free fit parameter. At the presently found level of accuracy, 
a fully equilibrated hadron gas (i.e. \gs=1) cannot be ruled out in all examined
collisions except in Pb--Pb, where \gs deviates from 1 by more than $4\sigma$.
This result does not agree with a recent similar analysis of Pb--Pb data 
\cite{heppe} imposing a full strangeness equilibrium. The main reason of this
discrepancy is to be found in the different data set used; whilst in 
ref.~\cite{heppe} measurements in different limited rapidity intervals have
been collected, we have used only  particle yields extrapolated to 
full phase space. The temperature values that we have found essentially agree 
with previous analyses in Au--Au collisions \cite{cor} and 
estimates 11.7 $A$ GeV \cite{stachel}.\\ 
The $T$ value in Pb--Pb is strongly affected by high mass particle measurements, 
such as $\phi$ and $\Xi$. A recent significant lowering of the $\Xi$ yield 
measured by NA49 \cite{barton} with respect to a previous measurement \cite{NA49-xi} 
results in a decrease of estimated temperature value from about 180 MeV to the
actual 160 MeV.

Forthcoming lower energy Pb--Pb and high energy Au--Au data at RHIC should allow 
to clarify the behaviour of strangeness production in heavy ion collision.

\section*{Acknowledgements} 
We are very grateful to N. Carrer, U. Heinz, M. Morando, C. Ogilvie 
for useful suggestions and discussions about the data. We especially 
thank H. Oeschler for his help with the GSI SIS data and R. Stock for
his help with NA49 data. 
\newpage

\newpage
\newpage

\end{document}